\documentstyle[12pt,aasms4]{article}

\lefthead{Barger et al.}

\begin{document}
\title{New Constraints on the Luminosity Evolution of Spheroidal Galaxies 
in Distant Clusters}
\author{A.~J.~Barger,\altaffilmark{1}
A.~Arag\'{o}n-Salamanca,\altaffilmark{2}
I.~Smail,\altaffilmark{3}
R.~S.~Ellis,\altaffilmark{2}
W.~J.~Couch,\altaffilmark{4}
A.~Dressler,\altaffilmark{5}
A.~Oemler,\altaffilmark{5}
B.~M.~Poggianti,\altaffilmark{2,}\altaffilmark{6} \and
R.~M.~Sharples\altaffilmark{3}
}

\altaffiltext{1}{Institute for Astronomy, University of Hawaii, 2680 Woodlawn Drive, Honolulu, Hawaii 96822, USA}
\altaffiltext{2}{Institute of Astronomy, University of Cambridge, Madingley Road, Cambridge CB3 0HA, UK}
\altaffiltext{3}{Department of Physics, University of Durham, South Road, Durham DH1 3LE, UK}
\altaffiltext{4}{School of Physics, University of New South Wales, Sydney 2052,
Australia}
\altaffiltext{5}{The Observatories of the
Carnegie Institution of Washington, 813 Santa Barbara St., Pasadena
CA 91101-1292, USA}
\altaffiltext{6}{Royal Greenwich Observatory, Madingley Road, Cambridge CB3 
OEZ, UK}


\begin{abstract}
We investigate various probes of luminosity evolution in the rich
cluster environment, concentrating in particular on the spheroidal
(E/S0) galaxies, using a newly-constructed catalog of
morphologically-classified {\it Hubble Space Telescope} {\it (HST)}
images of galaxies observed in the cores of 13 clusters with redshifts
$0.17\le z\le 0.56$. An important distinction of this study compared to
earlier work is the availability of new near-infrared ground-based
photometry for a substantial subset of our {\it HST} fields, which we
have used to select and study the various populations. We find no
significant change in the characteristic luminosity, $M_K^{\ast}$, of
the spheroidal populations at redshifts of $0.31$ and $0.56$.
As a more sensitive probe of
luminosity evolution, we analyze the {\it surface photometry} of our
{\it HST}-classified {\em ellipticals} by deriving effective metric
radii, $R_e$, and mean effective surface brightnesses,
$\langle\mu\rangle_e$. At the standard condition corresponding to $R_e
= 1\,$kpc, we find convincing evidence of evolutionary brightening in
both rest-frame $B$ and $K$ light, consistent with model predictions
based on the passive evolution of stellar populations.
\end{abstract}

\section{Introduction}

A remarkable empirical correlation of size, surface brightness, and
central velocity dispersion exists for elliptical galaxies
(\markcite{dress87}Dressler et al.\ 1987;
\markcite{dd87}Djorgovski \& Davis 1987). 
This {\em Fundamental Plane} relation implies important constraints
on the underlying dynamics of galaxy evolution. The virial
theorem provides a qualitative argument for a relationship between
these three observables (e.g.\ \markcite{djor93}Djorgovski \& Santiago 1993). 

Although extensive Fundamental Plane studies have been conducted at low
redshifts (e.g.\ \markcite{bender92}Bender, Burstein \& Faber 1992;
\markcite{guzman93}Guzm{\'a}n, Lucey \& Bower 1993;
\markcite{saglia93}Saglia, Bender \& Dressler 1993;
\markcite{pahre95}Pahre et al.\ 1995;
\markcite{jfk96}J\o rgensen, Franx \& Kj\ae rgaard 1996),
it is both difficult and expensive in telescope time to obtain 
the necessary spectroscopy at higher redshifts. Only recently, with
the advent of large telescopes, has it become
possible to make direct Fundamental Plane measurements
at $0.1<z<0.6$ (\markcite{vDF96}van Dokkum \& Franx 1996;
\markcite{jh97}J\o rgensen \& Hjorth 1997;
\markcite{pahre97}Pahre et al.\ 1997;
\markcite{kelson97}Kelson et al.\ 1997), but ongoing efforts
will soon make the Fundamental Plane a primary
diagnostic in determining the evolutionary history of early-type galaxies.

Fortunately, relatively tight correlations are also found in the 
{\em Kormendy Relation} (\markcite{korm77}Kormendy 1977), 
a projection of the Fundamental Plane
involving the effective radius, $r_e$, and the mean effective surface brightness
within that radius, $\langle\mu\rangle_e$. 
The advantage of the Kormendy projection is
that it can be explored with large data samples at high redshifts without the
need for spectroscopic observations. The low scatter about the Kormendy
Relation of $\langle\mu\rangle_e$ versus $R_e$, where $R_e$ is the effective 
metric radius, allows a reasonably accurate determination of 
$\langle\mu\rangle_e$ 
at a standard condition (SC), such as $R_e = 1\,$kpc. The dependence of
$\langle\mu\rangle_e^{SC}$ with redshift is determined 
by the $(1+z)^{-4}$ dimming due to the expansion of the Universe,
the $k$-corrections, passive
luminosity evolution, and dynamical evolution. The surface brightness
dimming is a firm theoretical prediction, stringently tested by the
observed Planck distribution of the cosmic background radiation
\markcite{mather94}(Mather et al.\ 1994). 
The $k$-corrections can be accurately deduced from local
spectral energy distributions; consequently, we can infer the
luminosity evolution with redshift and compare with predictions from
evolutionary synthesis models. 

Observational studies of color evolution in the early-type galaxy 
populations of distant clusters have demonstrated that changes with 
redshift are consistent with passively
evolving stellar populations formed at high redshift, $z_f>2$
(\markcite{alfonso93}Arag{\'o}n-Salamanca et al.\ 1993;
\markcite{rakos95}Rakos \& Schombert 1995;
\markcite{stan95}Stanford, Eisenhardt \& Dickinson 1995, 1998;
\markcite{oke96}Oke, Gunn \& Hoessel 1996;
\markcite{lubin96}Lubin 1996).
Similar constraints are inferred from the
small scatter and zero-points of the color-magnitude relations at both low
(\markcite{bower92}Bower, Lucey \& Ellis 1992)
and high redshifts (\markcite{dick97}Dickinson 1997;
\markcite{ellis97}Ellis et al.\ 1997;
\markcite{stan98}Stanford, Eisenhardt \& Dickinson 1995, 1998;
\markcite{kodama98}Kodama et al.\ 1998)
and from the observed evolution of the Mg\,b - $\sigma$ relation 
(\markcite{bzb96}Bender, Ziegler \& Bruzual 1996;
\markcite{bz97}Ziegler \& Bender 1997).

Recent observations with the {\it Hubble Space Telescope}
{\it (HST)} have made morphological studies of cluster galaxies feasible
(e.g.\ \markcite{dress94}Dressler et al.\ 1994;
\markcite{couch94}Couch et al.\ 1994). Since a Kormendy Relation 
analysis is based on size and surface brightness measurements, it
requires only high resolution imaging data
for a large number of cluster galaxies.
Several groups have recently conducted Kormendy Relation analyses to 
determine the amount of luminosity evolution in the cluster
elliptical galaxy population
(\markcite{pahre96}Pahre et al.\ 1996;
\markcite{bso96}Barrientos, Schade \& L\'opez-Cruz 1996;
\markcite{schade96}Schade et al.\ 1996;
\markcite{schade97}Schade, Barrientos \& L\'opez-Cruz 1997;
\markcite{dick97}Dickinson 1997). These authors generally find
that the luminosity evolution they detect is consistent with
or somewhat smaller than
expectations based on models of passively evolving old stellar 
populations. 

For evolutionary studies to be truly quantitative, large samples of
ellipticals from clusters over a wide range of redshifts are essential,
with the high redshift data being particularly important. In this paper
we analyze the luminosity evolution of a population of cluster galaxies
in the redshift range $0.17\le z\le 0.56$ using a large {\it HST}
cluster data set. Although there have been many recent studies of the
evolution of the luminosity function for field galaxies
(e.g.\ \markcite{rse97}Ellis 1997 and references therein),
very few studies have addressed the issue in higher-density environments.
An important feature of our analysis is the availability
of new and comprehensive near-infrared observations over the {\it HST}
fields for a large subset of the clusters in our sample. This has
enabled us to construct $K$-band luminosity functions for the various
cluster galaxy populations. $K$-band magnitudes are
less affected by $k$-correction uncertainties than optical magnitudes
and give a more robust indication of long-term changes in
the stellar populations. We then present new rest-frame $B$ and $K$-band 
measurements of surface photometry for the morphologically classified 
cluster {\em ellipticals}. From the intercept of the Kormendy Relation at
the standard condition $R_e=1\,$kpc, we measure the luminosity
evolution in our cluster elliptical galaxy populations between these
earlier epochs and the present.

A plan of the paper follows. Our observations and procedures are
discussed in \S 2. In \S 3 we present $K$-band luminosity functions for
the cluster galaxies both globally and by morphological type. We
discuss the extent to which our results for the spheroidal
population might be affected by the changing fractions of Es and S0s
with redshift. In \S 4 we do curve-of-growth analyses on our
morphologically selected cluster ellipticals to determine the relevant
parameters for a Kormendy Relation study. We then determine mean
effective surface brightness values at our standard condition to obtain
the luminosity evolution with redshift of the elliptical population in
the the rest-frame $B$-band. Finally, we couple our
optical-near-infrared colors with our $B$-band surface brightness
measurements to construct the Kormendy Relation in the $K$-band. This
provides a consistency check on the results of our $B$-band Kormendy
Relation analysis. In \S 5 we discuss our evolutionary results in the 
context of the Butcher-Oemler effect and the age of the stellar populations
in the elliptical galaxies. In \S 6 we summarize our conclusions.

\section{Observations and Reduction}

\subsection{{\it HST}/WFPC2 Observations}

The data set used in our analysis comprises 13 rich clusters with
redshifts between $z=0.17$ and $z=0.56$ imaged with the
Wide Field and Planetary Camera 2 (WFPC2) on board the {\it HST}. 
Eleven of these clusters
were observed as part of two studies of the morphologies of galaxies in
distant clusters, seven from the `MORPHS' project 
(\markcite{smailmorphs97}Smail et al.\ 1997a;
\markcite{ellis97}Ellis et al.\ 1997;
\markcite{dress97}Dressler et al.\ 1997)
and the remaining four from a project by 
\markcite{couch98}Couch et al.\ (1998).  
The two other clusters (Cl0024$+$16 and
A2390) were taken for lensing studies of these systems, and these images
were kindly provided to us by the project PIs (Profs.\ Turner and
Fort, respectively). Although these clusters were not selected through any
systematic criteria, they cover a range of a factor of three in richness and
represent many of the best examples of populous,
well-studied clusters at moderate redshifts. The spatial distributions
of the galaxies in the clusters, which vary from highly concentrated and 
regular to very chaotic, are evidence that the clusters cover a wide range in 
dynamical state. There is also an order-of-magnitude spread in their
masses and X-ray fluxes (\markcite{smailmass97}Smail et al.\ 1997b)

All 13 clusters have been the subject of long,
pointed observations in either the F702W or F814W filters. In addition,
all of the clusters observed in F814W have been imaged in the
F555W filter, with the exception of Cl0024$+$16 for which the additional
imaging was done in the F450W filter. A complete list of the clusters used
in this study are
given in Table~\ref{tab1}. These data were reduced in a standard
manner using the {\sc IRAF STSDAS} package and the {\sc CRREJ} stacking 
utility. Object catalogs for the final images were obtained using the
SExtractor package (\markcite{bertin96}Bertin \& Arnouts 1996), 
which provides positions and
photometry for all the objects detected. For our analysis we have
retained the WFPC2 color system and hence use synthetic zero points from
\markcite{holtzman95}Holtzman et al.\ (1995).
We refer to magnitudes in this system as $B_{450}$,
$V_{555}$, $R_{702}$, and $I_{814}$. More details of the reduction and
cataloging of the samples can be found in 
\markcite{couch98}Couch et al.\ (1998)
and \markcite{smailmorphs97}Smail et al.\ (1997a).

\placetable{tab1}

\subsection{New Near-Infrared Observations and Reduction}
\label{irobs}

In addition to the high resolution imaging with the {\it HST} of the clusters
discussed above, we have also obtained new near-infrared ($K$)
observations covering the same areas imaged by the {\it HST} for 6 of these
clusters. Table~\ref{tab2} contains a log of these infrared (IR) observations. 

Four clusters \ (Cl0016$+$16, \ Cl1601$+$42, 3C295 and Cl0024$+$16)
were imaged in the $K_s$ band in 1995 July 20--22 with a
near-infrared camera (\markcite{murphy95}Murphy et al.\ 1995) 
on the Palomar Observatory 1.52-m
telescope (P60) using a $256\times 256$ NICMOS-3 array. Individual clusters
were imaged with a layout of seven grids, each of which contained nine
different telescope pointings. The telescope was moved 10\,arcsec
between sets of grids and 15\,arcsec between
pointings. Each pointing was exposed for $6\times 40\,$s. Dark frames
were obtained at the beginning and end of each night with the same
exposure times and subtracted from the science images. The biweight
central location estimator\footnote{This estimator behaves like the
median when the number of data points is large but produces a better
S/N for relatively small samples 
(e.g.\ \markcite{beers90}Beers, Flynn \& Gebhardt 1990).}
of individual images taken over an $\simeq 36\,$min period
around a given observation produced a flat-field. The pixel scale for
these observations was $0.62\,$arcsec\,pixel$^{-1}$. Reductions were
completed using the {\sc FIGARO} package written by Keith Shortridge. All
nights were photometric, and the new {\it HST} faint standards were used for
calibration (Persson, private communication). Between three and four standards
were observed per night with a typical rms residual scatter less than 0.02\,mag.

The clusters Cl0054$-$27 and Cl0412$-$65 were imaged in 1994 November 11--12 in
$K^\prime$ with the IRIS infrared camera on the 3.9-m Anglo-Australian
Telescope (AAT) using a $128\times 128$ HgCdTe detector array. 
To cover the area imaged by the {\it HST}, a $2\times 2$ mosaic was
completed for each cluster using a layout of four grids, each containing
nine different telescope pointings. The telescope was moved 10\,arcsec between
pointings, and individual pointings were exposed for $12\times 10\,$s.
Dark frames with the same exposure times were obtained frequently and
subtracted from the images. The flat-field was obtained as above, using
an $\simeq 18\,$min period around a given observation. We artificially
divided each pixel into four to obtain a pixel scale of 
0.395\,arcsec\,pixel$^{-1}$, as described in 
\markcite{barger96}Barger et al.\ (1996). Both nights were
photometric and repeated observations of standard stars in the 
\markcite{allen92}Allen (1992)
list yielded photometry with errors $<0.04\,$mag rms.

Infrared-selected galaxy samples were automatically constructed using
the {\sc APM} software (\markcite{irwin85}Irwin 1985) in the 
Starlink {\sc PISA} implementation.
An object was detected in the images when four or more connected pixels
each had counts larger than 3$\sigma$ above the background. Integrated
photometry in a 5\,arcsec diameter aperture was obtained using the
{\sc PHOTOM} package in Starlink, which allows for a local estimation of
the sky. The size of the aperture was chosen large enough to include most
of the light for a majority of the galaxies but not so large
that contamination from nearby objects would be important.
Random photometric errors were determined empirically from the 
scatter estimated in sub-exposures and from the sky variance near
each object, since for the high $K$-band background this should be the
dominant source of error. Stars were rejected on the basis of their
morphologies on the {\it HST} frames. The limiting
magnitudes for the cluster $K$-band images, determined from the 
completeness of the number versus magnitude histograms,
are listed in Table~\ref{tab2}.

\placetable{tab2}

To obtain a homogeneous data set from these observations obtained
with different telescopes and instruments and using slightly different
photometric systems, we
converted all the near-infrared photometry to the standard $K$ system using 
the transformations $K=K^\prime+0.002-0.096(H-K)$ and
$K=K_s+0.009-0.124(H-K)$. The color terms were determined by
numerically folding the filter response curves with the near-infrared
galaxy spectral energy distributions (SED) described in
\markcite{alfonso93}Arag{\'o}n-Salamanca et al.\ (1993; hereafter AS93).
The size of this correction was always $\le0.08\,$mag. We will use the
data transformed to standard $K$ in all of our subsequent analyses.
The aperture photometry was also corrected for the effects of seeing 
(cf.\ AS93), but this is a
relatively unimportant effect ($\le0.1\,$mag) because of the large aperture
used (5\,arcsec diameter). 

Two of the clusters in the new $K$-band data set were previously imaged
by AS93 (Cl0054$-$27 and Cl0016$+$16). Although the new data presented
here are of better quality due to natural progress in near-IR detector
technology (typical uncertainties in AS93 photometry were 0.2 at
$18\,$mag, to be compared with $0.05$ in our new samples), the
photometry for the objects in common agrees well within the quoted
errors.

\section{The Near-Infrared Luminosity Function of Cluster Galaxies}

Galaxy luminosities are affected by evolutionary processes
occurring in the cluster galaxy populations, such as passive evolution
and the Butcher-Oemler effect. \markcite{smailmorphs97}Smail et
al.\ (1997a) examined these effects in our sample 
by studying the optical luminosity functions by morphology. In
this paper we examine the effects using the clusters' $K$-band
luminosity functions. Since the near-IR light is dominated by
near-solar-mass stars, $K$-band observations provide a good estimate of
the total stellar mass in old populations. Furthermore, near-IR
luminosities are considerably less affected by the occurrence of small,
star-forming bursts in galaxies than are the corresponding optical
luminosities (e.g.\ \markcite{barger96}Barger et al.\ 1996).
Consequently, $K$-band luminosity functions by morphology give a good
indication of the total fraction of old stars in galaxies of different
morphological types.

\subsection{The Global Luminosity Function}
\label{globalLF}

AS93 studied the $K$-band luminosity functions (LFs) of 11 clusters in
the $0.37\le z\le 0.92$ range and found no strong evidence for any
luminosity evolution. However, since the area coverage and depth of
their images provided only 10--20 galaxies per cluster, their
uncertainties in the determination of $M_K^\ast$ (the characteristic
luminosity) were of the order of the expected effect for galaxies
evolving passively; thus, no strong conclusions could be drawn. Recently,
\markcite{barger96}Barger et al.\ (1996) published a detailed study of
three $z\simeq 0.31$ clusters. They presented the combined $K$-band LF
for cluster galaxies spanning $\simeq 4.5\,$mag and 
compared it with the local field $K$-band LF of
\markcite{mobasher93}Mobasher et al.\ (1993). They concluded that the
LF for the three clusters had a flat faint-end slope ($\alpha\simeq
-1.0^{+0.13}_{-0.11}$) and a characteristic luminosity
similar to that of present-day field galaxies. No sign of strong
luminosity evolution was found, although the main
uncertainty was the lack of a measured $K$-band LF for present-day
clusters.

Using the $K$-band images presented in \S\ref{irobs} in combination
with the $K$-band images of the three clusters studied by
\markcite{barger96}Barger et al.\ (1996) (AC103, AC114 and AC118 at
$z=0.31$) and the cluster Abell~370 ($z=0.37$) studied by
\markcite{aes91}Arag{\'o}n-Salamanca, Ellis \& Sharples (1991), we
analyze the $K$-band galaxy LFs of 10 clusters in the $0.31<z<0.56$
range. The new photometry is complete to the limits specified in
Table~\ref{tab2}. The Abell~370 photometry is complete to $K=17.5$,
and the AC103, AC114 and AC118 photometry to $K=19.0$, although here we
only present data to $K=18.0$ for consistency with the other clusters.
The combined sample contains 672 galaxies.

P.~Eisenhardt and N.~Trentham (private communications) find that
local cluster $K$-band LFs (for Coma and for A1795 + A665 + A963, respectively)
are consistent within the errors with the local field $K$-band LFs of
\markcite{gardner97}Gardner et al.\ (1997) and
\markcite{mobasher93}Mobasher et al.\ (1993). Because we can directly
compare our magnitudes to those of Mobasher et al.\ (1993) 
(see Barger et al.\ 1996 for details)
after applying only small aperture corrections to our data
($0.08\,$mag at $z=0.31$ to $0.24\,$mag at $z=0.56$; cf.\ AS93),
we have decided to adopt the Mobasher et al.\
local field LF as the low-$z$ comparison sample for our current study.

Since not all of the galaxies in our $K$-selected samples have redshifts,
we use published $K$ counts by
\markcite{gardner94}Gardner, Cowie \& Wainscoat (1994)
to estimate the field contamination of our cluster data. 
These authors found that the field-to-field variations in their number counts 
were consistent with Poisson statistics. If the same holds true for our data, 
then the uncertainty introduced by the field subtraction has a very small
effect in the determination of $M_K^\ast$ in the Schechter function 
(\markcite{schec76}Schechter 1976). A $1\sigma$ variation in
the field number counts changes $M_K^\ast$ by much less than $1\sigma$ 
in the individual cluster luminosity functions. This uncertainty is
reduced even more when several cluster fields are binned together.

We find that the observed LFs for the individual clusters are visually
compatible with each other; thus, we improve the statistics by grouping
the clusters into three redshift bins: AC103, AC114 and AC118 at
$\langle z\rangle = 0.31$; Abell~370, Cl0024$+$16 and 3C295 at $\langle
z\rangle = 0.40$; and Cl0412$-$65, Cl1601$+$43, Cl0016$+$16 and
Cl0054$-$27 at $\langle z\rangle = 0.54$.

Since most galaxies, regardless of galaxy spectral type, have a similar
spectral energy distribution in the near-IR, the $k$-corrections are
expected to be independent of morphology
(e.g.\ \markcite{alfonsothesis91}Arag{\'o}n-Salamanca 1991;
\markcite{alfonso94}Arag{\'o}n-Salamanca et al.\ 1994). By folding the
$K$-band response function with the local SEDs of AS93 at the
appropriate redshifts, we find $k$-corrections which range from
$-0.36\,$mag at $z\simeq0.17$ to $-0.56\,$mag at $z\simeq0.56$. From
comparing different galaxy SEDs, we estimate that the uncertainties in
the $k$-corrections are $\leq 0.1$ at these redshifts.

Figure~\ref{bin} shows binned cluster galaxy LFs for each redshift group,
with the best fitting Schechter functions for a fixed faint-end slope of
$\alpha=-1$ drawn as solid lines. Note that the magnitude
range covered by most of our cluster data ($3-3.5\,$mag) does not allow
us to determine $\alpha$ with any confidence. Since 
\markcite{barger96}Barger et al.\ (1996) obtained
$\alpha=-1.0^{+0.13}_{-0.11}$ using a larger magnitude range for the $\langle
z\rangle = 0.31$ clusters, we have adopted $\alpha=-1.0$ for this study.

\placefigure{bin}

The LFs in the three redshift groups are very similar, and, in particular,
their $M_K^\ast$ values agree at the $\pm 1\sigma$ level.
Figure~\ref{morph}a (top) shows the combined LF for all the clusters. The
average $M_K^\ast=-25.4\pm0.1$ is in good agreement with the values
obtained by AS93 in the $0.37\le z\le0.92$ range and are only slightly
brighter than the value found by \markcite{mobasher93}Mobasher et al.\ (1993)
for field galaxies at $z\sim0$ ($M_K^\ast=-25.1\pm0.3$).
{\it Thus, over the limited redshift range where the cluster LF can be
studied uniformly, we cannot detect significant luminosity evolution.} 

\subsection{The Luminosity Function by Morphological Type}

In the previous section we combined the early- and late-type
populations in order to examine the evolution in the global luminosity
function. However, by combining disparate populations we may dilute the
actual evolution that may be occurring for individual classes. For the
early-types we expect mainly passive evolution, modified perhaps by
whatever physical process produces the Butcher-Oemler effect. For the
spirals, whose disks form stars over longer timescales at an
approximately constant star formation rate, we expect compensating
effects from a passively evolving old stellar population and from a new
population of young stars, the details of which depend on the specific 
star formation history.

Since we have morphological information for the cluster galaxies from
our {\it HST} imaging (\markcite{couch98}Couch et al.\ 1998;
\markcite{smailmorphs97}Smail et al.\ 1997a), we can examine this point
by studying the $K$-band LFs by morphological type. We divide
the galaxy samples into early-type (Es and S0s) and late-type (spirals
and irregulars) categories. The galaxy sample is partially reduced
because the area imaged by the {\it HST} is slightly smaller than that of
the ground-based near-infrared images.  

The question arises as to whether the interpretation of 
an evolutionary signal determined from a combined sample of Es and S0s 
may be confused because of the different evolution for the two sub-classes,
particularly given the significant decline in the S0 fraction with increasing
redshift seen by \markcite{dress97}Dressler et al.\ (1997). The point 
is an important one since \markcite{smailmorphs97}Smail et al.\ (1997a) 
note that separating face-on S0s and ellipticals is quite difficult even 
in {\it HST} images (many of the early-type galaxies in the Smail et al.\ 
catalogs are simply classified as E/S0 or S0/E). However, the paucity of S0s
in intermediate-redshift clusters (about 15\ per cent, cf.\
\markcite{smailmorphs97}Smail et al.\ 1997a; 
\markcite{dress97}Dressler et al.\ 1997), means we do not expect them to 
have a large effect on the combined 
LF (and, in particular, on $M_K^\ast$).  Nonetheless, we demonstrate
the effect quantitatively below.

In this analysis the field contamination is more uncertain than was the
case in \S 3.1 since reliable $K$-band number counts as a function of
morphology are not yet available. Instead, we use the {\it HST} Medium
Deep Survey counts in the $I$-band as a function of morphology
(\markcite{glaze95}Glazebrook et al.\ 1995) to estimate the counts in
$K$ for the different types. At the typical depth of our $K$-band
images ($K\simeq 18.5$) the median redshift is $z=0.5$
(\markcite{cowie96}Cowie et al.\ 1996).  At this redshift the average
color of early-type galaxies is $(I-K)=3.0$ and that of late-types is
$(I-K)=2.5$ (cf.\ AS93). Using these colors to convert the $I$-band
number counts as a function of morphology into $K$-band number counts
as a function of morphology, we predict that the field contamination is
about a factor~2 higher for the spiral and irregular population than
it is for the early-type population.

This procedure is not completely satisfactory since the input number
counts are $I$-band selected instead of $K$-band selected and may 
therefore contain a different mix of galaxy types at a given magnitude 
than would a $K$-band selected sample. 
Moreover, although using average colors at a given magnitude
will give the correct normalization of the $K$ counts at the typical
$K$ magnitude, the slope may not be correct. Given the
narrow range in magnitudes considered here, however, this should be a
second-order effect. Nevertheless, we recognize that significant
uncertainties could be present in this analysis, but without $K$ counts
as a function of morphology, all we can do is to try and quantify these
uncertainties. As an extreme case, let us assume that field
contamination affects both the early-types and the late-types {\it equally\/}
such that the field contamination of the E/S0s would be overestimated while 
that of the spiral/irregulars would be underestimated. In this scenario we derive
$M^\ast$ values for both the E/S0s and the later types which
deviate by slightly less than $1\sigma$ from the results
obtained using our adopted field-subtraction procedure. Thus, the
systematic uncertainties introduced by our procedure are likely of
the same order of magnitude as the estimated random errors.

We divide the early-type sample into two redshift bins at $z=0.34$ and
$z=0.52$. The best-fit Schechter function for both bins is
$M_K^\ast=-25.4\pm0.3$. Unfortunately, the uncertainty on $M_K^\ast$ is
quite large due to the small sample size. Considering that the expected
brightening of the stellar populations from passive evolution in this
redshift interval is $\sim0.3$, our luminosity function analysis is
insensitive to such evolution. This is consistent with the small
brightening in the rest-frame $V$-band ($\delta M_V^\ast\sim -0.3$)
found by Smail et al.\ (1997a) for the elliptical galaxies in the same
redshift range. 

\placefigure{morph}

In Figure~\ref{morph}a we show cluster LFs sub-divided by morphological
type and for the combined samples (cf.\ \S3.1) at the mean redshift
$z=0.43$.  The dominance in the near-IR of the early-type galaxies (66
per cent) over the late-types (34 per cent) in the cluster cores
($\simeq1\,$Mpc diameter) is clearly evident. Using the local
morphology-density relation (\markcite{dressmd80}Dressler 1980) at an
average projected galaxy surface density of
$\simeq30\,$galaxies\,Mpc$^{-2}$ (corresponding to the one we measure
in the {\it HST} images) we would expect about 18~per cent spiral and
irregular galaxies. Thus, we find that the fraction of late-types is
about a factor~2 higher in our sample than in low redshift clusters.
\markcite{dress97}Dressler et al.\ (1997) showed that the optical
morphological mix in these clusters changes dramatically with redshift.
In our data we see that this change in morphological mix is also
evident when the galaxies are selected in the near-IR, where the galaxy
selection is done more closely according to their total stellar mass.

In Figure~\ref{morph}b we return to the issue of E and S0 confusion and
examine the luminosity distribution of definite S0s in relation to the
LF of the combined E+S0s and to the LF with the S0s removed. Because the
field counts are not available for S0s, it is not possible to correct
the distribution for field contamination, but correction would only
strengthen the conclusions which follow. The number of S0s in our
sample is too small to shed much light on the strong decline with
redshift of the S0 galaxy fraction in rich clusters that
\markcite{dress97}Dressler et al.\ (1997) found. However, the influence
of S0 evolution on the results derived for the Es is likely to be quite
small, even allowing for some confusion between the two classes. The derived
$M_K^\ast$ changes by only $\simeq0.5\sigma$ (0.13 mag) when we exclude the 
S0 population. In summary, the minimal evolution of the E+S0 population
faithfully tracks the trend found for the pure elliptical population.

The slow evolution in the $K$-band LF of elliptical galaxies in rich
clusters contrasts with the recent claim by \markcite{KCW96}Kauffmann,
Charlot \& White (1996) that the population of field ellipticals has
evolved very rapidly since $z=1$. However, it must be remembered that
the environments to which these results apply are very different. In
the case of the field elliptical population, the strong evolution is
largely in the {\it number} of passively evolving red systems: a result
Kauffmann et al.\ interpret as indicative of the gradual formation of
field ellipticals from frequent galaxy mergers as predicted in the
hierarchical clustering theory. The color and luminosity evolution of
the oldest examples might well give similar results to the trends
located here so the contrast is largely in a {\it range of formation
ages} after which passive evolution remains a valid approximation to
the luminosity evolution.  In fact, the same hierarchical models (see
\markcite{KC97}Kauffmann \& Charlot 1997 and references therein)
predict a much slower evolution up to $z\simeq1$ for ellipticals in
rich clusters since in these very high density environments the
associated merging occurs very early, i.e., well beyond $z\sim1$ for
rich clusters observed at $z\sim0.5$.

Finally, we note that Fig.~\ref{morph}a shows that $M_K^\ast$ is
somewhat brighter for the E and S0 sample than for the late-type
galaxies. This reflects the fact that the brightest cluster
galaxies are usually early-type galaxies as is the case in nearby 
clusters (e.g.\ \markcite{binggeli88}Binggeli, Sandage \& Tammann 1988). 
The fact that the brightest luminosity bin in the LFs contains $\sim 2$ 
spiral galaxies (with a large error bar) is probably due to the 
larger field contamination for the later types.

In this section we have explored the near-IR LFs of cluster galaxy
populations both as a whole and divided into early- and late-type
morphologies. Due to the current lack of a suitable $z=0$ near-IR
cluster LF for use as a comparison template, we have been unable to
draw any strong conclusions concerning the luminosity evolution of
cluster galaxies. However, large area near-IR cluster surveys now
being carried out (e.g.\ Eisenhardt et al., in preparation) will soon
permit direct comparisons to be made. Nonetheless, we demonstrate that
any evolution consistent with our data must be less than or equal to
the expected amount for passive evolution over the redshift range
sampled. In the following sections we will concentrate on the cluster
{\it elliptical} population. The presence of the Fundamental Plane
(hereafter, FP) for such galaxies permits a more precise measurement of
the luminosity evolution in our sample.

\section{Evolution in Surface Brightness: the Kormendy Relation}

As a more sensitive probe of the luminosity evolution of the cluster elliptical
galaxy population, we now do a Kormendy Relation (KR) analysis. 
We select the ellipticals (T type $-$5) for this study from the visual 
morphological classification lists presented and described in
\markcite{couch98}Couch et al.\ (1998) and 
\markcite{smailmorphs97}Smail et al.\ (1997a). 
The cluster Abell~2390 is the only 
one not included in either of the above studies; we classify it as well
using the visual scheme. The classification lists are limited in 
magnitude to $R_{702}\leq23.5$ and $I_{814}\leq23.0$, as beyond these limits
the visual typings become significantly incomplete.
The reliability of the morphological selection of spheroidal galaxies 
in clusters at $z\sim0.55$ was examined in detail by 
\markcite{ellis97}Ellis et al.\ (1997). They
found that the visual classifications were robust to $I_{814}\leq21.0$
but over the interval $I_{814}=21.0-22.0$ the distinction between 
the early-type classes became increasingly uncertain. 

In order to reduce contamination of the cluster elliptical samples by
foreground/back{-}ground field ellipticals, we have trimmed our samples
using all available color information and spectroscopy
(\markcite{dg92}Dressler \& Gunn 1992;
\markcite{couch98}Couch et al.\ 1998). The color criteria used to
isolate the ellipticals which lie on the early-type galaxy locus,
and hence to distinguish cluster members, are sufficiently
liberal that cluster members will not be removed
($\pm0.3\,$mag around the color-magnitude relation in all colors).
Only a few galaxies are removed from the clusters by these cuts.

Two previous high-redshift KR studies
(\markcite{pahre96}Pahre et al.\ 1996; 
\markcite{schade97}Schade et al.\ 1997) adopted
a redshift-independent limiting absolute magnitude cut for the 
cluster galaxy samples. 
However, we note that if the luminosities of cluster 
ellipticals evolve passively in the sense that at higher redshifts the 
galaxies become brighter, then by adopting a common absolute magnitude limit, 
the true luminosity evolution of a given sample of
cluster ellipticals will necessarily be underestimated. 
That is, galaxies which were
below a given magnitude cut-off at a lower redshift would brighten
as the redshift increased and hence might rise above the fixed cut-off at 
a higher redshift, thereby contaminating the sample.

A consistent treatment of a passively evolving elliptical population in a
KR analysis therefore requires that the evolution of the limiting 
absolute magnitude 
{\it also} be taken into account. In principle this requires an 
iterative procedure; in order to impose an absolute magnitude cut 
the amount of luminosity evolution needs to be known, but this is the 
quantity we are trying to determine. As discussed in \S~1,
both the color evolution of the elliptical population and the small scatter
observed in their $U-V$ colors imply old ages for the stellar populations
($z_f>2$); thus, for consistency with such results, we assume $z_f>2$ 
in our analysis. To predict the luminosity evolution of a passively evolving 
elliptical galaxy at a given redshift, we use the 1996 version
of the \markcite{bc96}Bruzual \& Charlot (hereafter, BC96) stellar 
evolutionary models. In practice we find that an iterative procedure is
not really necessary since the overall comparisons of the data with 
the models do not depend very much on the choice of $z_f$ and our 
sliding magnitude cut gives closely similar data points for any $z_f$. 
We stress, however, that the optical data points are
very dependent on the non-constancy of the absolute magnitude cut, as will
be demonstrated subsequently.

We determine the two KR parameters (the half-light radius, $r_e$, and the mean 
surface brightness interior to that radius, $\langle\mu\rangle_e$) by fitting 
the analytic de Vaucouleurs law to growth curves constructed from aperture
photometry of the morphologically classified ellipticals in the 
{\it HST} F702W and F814W frames.
Galaxies for which the growth curves are not well fit by the de Vaucouleurs 
law ($\sim20$\ per cent), presumably because they have exponential components 
to their light profiles, are rejected. 
We correct all $\langle\mu\rangle_e$ measurements
for the observationally established 
$(1+z)^{-4}$ surface brightness dimming by subtracting $10\log (1+z)$. 

For the clusters for which we have near-infrared imaging, we convert the 
{\it HST} $\langle\mu\rangle_e$ measurements into $\langle\mu_K\rangle_e$ 
using the equations $\langle\mu_R\rangle_e-\langle\mu_K\rangle_e=(R_{702}-K)$ 
and $\langle{\mu_I}_{814}\rangle_e-\langle\mu_K\rangle_e=(I_{814}-K)$. 
The $(R_{702}-K)$ and $(I_{814}-K)$ colors are based on measurements within
$5\,$arcsec diameter circular apertures,
corrected for mean aperture color effects using $\Delta(V-K)/\Delta
\log (2\times r_e/5)=-0.1$ (\markcite{pfa79}Persson, Frogel \& Aaronson 1979),
where $r_e$ has units of arcsec. We did not first convolve
the {\it HST} images to the resolution of the ground-based
near-IR imaging; however, the difference in magnitude 
between a convolved and an unconvolved image is negligible. (The difference
between an unconvolved image and one convolved to 1\,arcsec seeing is only
0.01-0.02\,mag, in the sense that the unconvolved magnitude is brighter. When 
the image is convolved to 1.5\,arcsec resolution, the difference is
0.03-0.04\,mag.) Since the fields are relatively crowded, we occasionally
are unable to use an object selected in the infrared because it
corresponds to several {\it HST} objects.

We correct all $\langle\mu\rangle_e$ values for foreground Galactic
extinction using E(B-V) values (Table~\ref{tab1}) obtained from the 
\markcite{bh82}Burstein \& Heiles (1982) maps.
The $k$-corrections in $K$ are well determined (\S\ref{globalLF}), and
the uncertainties associated with optical $k$-corrections can be minimized
by reducing the results to rest-frame $B$ because the 
F814W and F702W passbands closly approximate this band at the higher
redshifts sampled.
The latter corrections are determined by folding the {\it HST}
filter response functions with the local SEDs from AS93 at the appropriate 
redshifts. 

Figure~\ref{figBkorm} shows
$\langle\mu_B\rangle_e$, corrected for cosmological surface brightness 
dimming, versus $\log R_e$ ($H_o=50\,$km s$^{-1}\,$Mpc$^{-1}$, $q_o=0.5$) 
for the 13 {\it HST} clusters, grouped into redshift bins. The
sliding absolute magnitude cut is based on $z_f=5$ and is illustrated 
in each of the frames of Fig.~\ref{figBkorm} by
the dotted line. We adopt the value $M_B=-19.83$ for the absolute
magnitude cut locally, since
that is the completeness limit of the local $B$-band data of
\markcite{schade97}Schade et al.\ (1997).
Figure~\ref{figKkorm} similarly shows $\langle\mu_K\rangle_e$,
corrected for cosmological surface brightness dimming, 
versus $\log R_e$ for the nine
clusters for which we have near-infrared data. Once again, the sliding
absolute magnitude cut, based on $z_f=5$, is indicated in each of the 
frames of Fig.~\ref{figKkorm} by a dotted line. 
This time we have adopted the value $M_K=-22.93$ for the local absolute 
magnitude cut since it approximately matches the limit of the local $K$-band
data of \markcite{pahre95}Pahre et al.\ (1995).

\placefigure{figBkorm}

\placefigure{figKkorm}

In order to deduce the evolution with redshift of the elliptical populations 
in clusters, we choose to employ a standard condition
to compare galaxies of the same physical size at different redshifts.
The full FP relation is
\begin{eqnarray}
\langle\mu\rangle_e = -a\log\sigma + b\log R_e + c
\end{eqnarray}
We condense the information contained in the rest-frame
Kormendy plots in each passband and for each cluster to a single
number by defining the quantity
\begin{eqnarray}
\mu' &\equiv& \langle\mu\rangle_e - b\log R_e \label{eq:mu'equiv}
\end{eqnarray}
which effectively translates all of the $\langle\mu\rangle_e$
data points to $\log R_e = 0$, the value of $R_e$ in kpc that we choose 
as our standard condition.
Here $b$ is the slope parameter of the full FP relation.

We adopt the FP slopes from the local $B$-band study of 
\markcite{jfk96}J\o rgensen et al.\ (1996) and the local
$K$-band study of \markcite{pahre95}Pahre et al.\ (1995).
The slope is found to be $b=3.0$ in the case of J\o rgensen et al.\ and
$b=3.2$ in the case of Pahre et al., both of which are
compatible with our data; least-square fits to the data {\em brighter} than the
absolute magnitude cut for these slopes are illustrated by the solid lines
in Figs.~\ref{figBkorm} and ~\ref{figKkorm}.

We determine the error on $\mu'$ from the equation
\begin{equation}
\delta\mu'=\biggl[{1 \over {N(N-1)}}\sum(\mu'-\overline{\mu'})^2\biggr]^{1/2}
\end{equation}
where $N$ is the number of data points.
The intrinsic scatter in the Kormendy Relation arises from the 
neglect of the $\log\sigma$ term in the FP relation 
and dominates the statistical uncertainties in
our de Vaucouleurs parameter determinations. It is difficult to
ascertain statistical uncertainties when using integrated growth curves, but
from direct de Vaucouleurs fits to elliptical 
isophotes, we find the statistical 
uncertainties in $\langle\mu\rangle_e$ to be an order of magnitude smaller
than the intrinsic scatter uncertainties.

For comparison with our high redshift results, we have deduced
local cluster data points from the
optical data of \markcite{schade97}Schade et al.\ (1997) and from
the near-infrared data of \markcite{pahre95}Pahre et al.\ (1995), although
it should be cautioned that different techniques
were used by these observers in determining $\langle\mu\rangle_e$ and $R_e$.
The values of $\Delta{\mu'}$ for the two passbands, defined as 
\begin{equation}
\Delta{\mu'}=\mu'_z - \mu'_{local}
\end{equation}
are plotted in Figs.~\ref{Bsc}, binned in redshift as 
in Figs.~\ref{figBkorm} and \ref{figKkorm}, versus $\log(1+z)$.
We use different symbols for each of the data points for clarity.

\placefigure{Bsc}

In Figure~\ref{Bsc} we plot c-model results from BC96 to compare
the data with passive evolution expectations. The curves
assume a \markcite{salpeter55}Salpeter (1995) initial mass function,  
solar metallicity, and a 1~Gyr initial burst at the redshift of galaxy 
formation, $z_f$. The model predictions
involve an additive normalization constant, which we have adjusted to match 
the central values of the local data points.
We assume for all of our figures a $q_o=0.5$ cosmology with
$H_o=50\,$km s$^{-1}\,$Mpc$^{-1}$.

Figure~\ref{Bsc}a illustrates the effects of applying a constant 
absolute magntitude cut, $M=-19.83$, to the optical data. Figure~\ref{Bsc}b
shows the corresponding results when a sliding magnitude cut is used
($M=-19.83$ locally). It is apparent that we would
underestimate the amount of luminosity evolution occurring in the elliptical
galaxy population if we were to rely on a constant absolute magnitude cut. 
Although we have shown the BC96 passive evolutionary models for both $z_f=2$ 
and $z_f=5$ in Fig.~\ref{Bsc}b, we have only used $z_f=5$ in applying
the absolute magnitude cut to the data.
However, we find that the data values are quite insensitive to both the value 
of $z_f$ assumed and the cosmology.

Figure~\ref{Bsc}c shows the $z_f=5$ and $z_f=2$ BC96 near-infrared 
curves plotted on our complete near-infrared data sample. Since at present
it has only been possible for us to carry out a KR test in the $K$-band using 
surface brightness and scale-length measurements made in the optical,
the results should be viewed only as verification of our more accurate $B$-band 
analysis. $K$-band surface brightness analyses will most
certainly be improved upon in the near future when IR KR parameters can be
measured directly from high resolution {\it HST} NICMOS observations.

We see from these figures that both the optical $\mu'_B$ and the infrared  
$\mu'_K$ data are compatible with the model passive evolution curves. 
Our $B$-band results are in good agreement with those of 
Schade et al.\ (1997), once the sliding absolute magnitude cut has been 
imposed on their data. 
We note that the data point containing the three clusters at $z=0.31$ appears 
to be somewhat greater than $2\sigma$ above the model lines
(each of the three clusters contained in this point were found to be 
equally discrepant). We do not have an explanation for why this should be 
the case.

\section{Is Apparent Passive Evolution Consistent with the Butcher-Oemler Effect?}

There is compelling evidence from cluster data at moderate
redshifts for significant starburst activity, some component of
which is likely to be occurring in
the elliptical population (Butcher-Oemler effect; 
\markcite{dg82}Dressler \& Gunn 1992;
\markcite{cs87}Couch \& Sharples 1987;
\markcite{bianca96}Poggianti \& Barbaro 1996;
\markcite{couch98}Couch et al.\ 1998).
An analysis by \markcite{barger96}Barger et al.\ (1996)
compared observed number distributions of star-forming
galaxies in three clusters at $z=0.31$ with corresponding model
expectations. The models assumed that a population of cluster galaxies
had undergone minor secondary bursts of star-forming activity (creating
10--20 per cent of the total stellar mass) within the past $\sim
2\,$Gyr. These authors estimated that $\sim30$ per cent of the total cluster
population (including 30 per cent of early-types) needed to be
involved in such activity for the models to adequately reproduce the
observations, although this does not necessarily imply that all of the 
observed bursts are the result of major merging events (Couch et al.\ 1998). 

Since such a high fraction of bursting galaxies might be expected to
increase the average luminosity of the elliptical galaxy population,
we think it appropriate to consider here how much the {\it average} 
surface brightness of the elliptical population might
increase over the amount predicted by passive evolution models in order
to test whether there is a contradiction between our current results 
(in which the observed luminosity evolution is seen to be compatible with 
passive) and those of previous Butcher-Oemler studies.

We use the models of \markcite{barger96}Barger et al.\ (1996) to
estimate the overall statistical brightening that would result 
if 30 per cent of a cluster elliptical population at $z=0.31$ had experienced 
10 per cent secondary bursts within the past $\sim 2\,$Gyr.
For an entirely elliptical galaxy population at $z=0.31$,
we find that the average brightening is only 
$\delta\langle\mu_B\rangle_e=0.16\,$mag arcsec$^{-2}$ in $B$ 
and $\delta\langle\mu_K\rangle_e=0.09\,$mag arcsec$^{-2}$ in $K$. Since 
this amount is relatively insignificant at this redshift,
we conclude that the brightening due to minor starbursts in some
proportion of the elliptical 
population would not cause an appreciable shift in the passive 
evolution curves and that our Kormendy Relation results are
unlikely to be affected significantly by starburst phenomena. 

\section{Summary}

We analyzed {\it HST} data for 13 rich clusters over the redshift
range $z=0.17$ to $z=0.56$ and extended the database by securing new
near-infrared images over the same fields for a substantial subset.
Taking advantage of the benefits of near-infrared selection for studies
of galaxy evolution over a range in redshift, we determined
the total IR LFs for our clusters, as well as the LFs by morphological 
type. Restricting the sample to {\it bona fide} cluster
ellipticals, we then analyzed the evolution with redshift of the
Kormendy Relation of mean effective surface brightness,
$\langle\mu\rangle_e$, versus effective metric radius, $R_e$, in both
the $B$ and $K$ bands. Our principal results are as follows:

$\bullet$ The total cluster and pure elliptical LFs
do not show any evidence for luminosity evolution over
the redshift range sampled. This result is consistent with passive
evolution in the elliptical population, although quantitative estimates
are limited by the absence of a reliable local cluster elliptical LF. 
We demonstrate that
our result is not affected by possible contamination of the pure
elliptical sample from misclassified
S0s following a different evolutionary history. 

$\bullet$ The LFs as a function of morphology further indicate that the
galaxy population is dominated by early-types (64 percent) and that the
fraction of spirals and irregulars is about a factor of two larger than
the fraction observed in similar dense environments at low $z$. The
data also suggest that the characteristic luminosity of the early-type
galaxies is brighter by $\sim 0.5\,$mag than the characteristic
luminosity of the later types.

$\bullet$ A sliding magnitude cut with redshift needs to be implemented 
in Kormendy Relation analyses.

$\bullet$ The surface brightness 
${\mu^\prime}=\langle\mu\rangle_e-b\log R_e$ values
for the high redshift cluster samples at the standard condition
of mean effective radius $R_e=1\,$kpc are in accord with passive evolution
expectations of the BC96 code in both the $B$ and $K$ bands.

\begin{acknowledgements}

We wish to thank Ray Lucas at STScI for his enthusiastic and able
help in the efficient acquisition of these {\it HST} observations. We also
thank Roger Davies, Paul Hewett, Bob Abraham, and Jarle Brinchmann for useful
discussions and suggestions, and William Vacca for invaluable assistance.
Finally, we thank the referee, Mark Dickinson, and the editor, Greg Bothun,
for suggestions and comments which substantially improved the paper.
AJB acknowledges support from the Marshall Aid Commemoration Commission 
and from NASA through contract number P423274 from the University of Arizona,
under NASA grant NAG5-3042. AAS acknowledges support
from the Royal Society, IS, RSE and RMS from the 
Particle Physics and Astronomy Research Council, and
BMP from the Formation and Evolution of Galaxies network set up by the European
Commission under contract ERB FMRX-CT96-086 of its TMR program.
WJC acknowledges support
from the Australian Department of Industry, Science and Technology, the 
Australian Research Council, and Sun Microsystems.
AD and AO acknowledge support from NASA through STScI grant 3857.

\end{acknowledgements}

\newpage
 
\figcaption
{Absolute magnitude distribution for the combined 
sample of cluster galaxies in three redshift groups. Error
bars are based on Poisson statistics, and the smooth lines show the 
best-fitting Schechter luminosity functions. The vertical scale has been
shifted by 1 logarithmic unit for the $\langle z\rangle = 0.4$ clusters
and by 2 units for the  $\langle z\rangle = 0.54$ clusters for clarity.
\label{bin}}

\figcaption
{(a) The luminosity function by morphological type 
for the combined cluster galaxy sample at the mean redshift z=0.43. 
Open circles: ellipticals and S0s; open triangles: spirals
and irregulars; filled dots: all galaxies. The error bars are based on
Poisson statistics, and the smooth lines show the best-fitting Schechter
luminosity functions. (b) The LF for the combined elliptical and S0 sample
compared to that determined with the definite S0s removed. The luminosity
distribution of S0s is also shown, uncorrected for
field contamination. \label{morph}}

\figcaption
{Transformed mean effective surface brightness,
$\langle\mu_B\rangle_e$, corrected for
$-10\log(1+z)$ cosmological surface brightness dimming, versus $\log R_e$
($q_o=0.5$, $H_o=50$) for the 13 {\it HST} clusters in rest-frame $B$.
The panels show clusters grouped in redshift as follows: (a) Cl1601$-$42,
Cl0016+16, and Cl0054$-$27 ($\bar z=0.55$); 
(b) 3C295 and Cl0412$-$65 ($\bar z=0.48$);
(c) Cl1447$+$23, Cl0024$+$16, Cl0303$+$17 ($\bar z=0.39$);
(d) AC103, AC114, and AC118 ($\bar z=0.31$);
(e) A2390 ($z=0.23$);
and (f) A2218 ($z=0.17$).
The dotted lines represent the sliding absolute magnitude cut (see text)
applied for $z_f=5$ ($q_o=0.5, H_o=50$), assuming $M_B = -19.83\,$mag locally.
The solid circles are galaxies whose magnitudes lie
above the magnitude cut, and the open circles represent data falling below 
the cut. Least-squares fits to the data above the absolute magnitude cut,
using the slope of a local full Fundamental Plane study by
J\o rgensen et al.\ (1996), $b=3.0$, are indicated on the plots as solid lines.
\label{figBkorm}}

\figcaption
{$\langle\mu_K\rangle_e$ versus $\log R_e$ for the nine
clusters observed in $K$, corrected
for cosmological surface brightness dimming and $k$-corrections.
The panels show clusters in the redshift groupings: (a) Cl1601$+$42,
Cl0016$+$16, and Cl0054$-$27 ($\bar z=0.55$); 
(b) 3C295 and Cl0412$-$65 ($\bar z=0.48$); 
(c) Cl0024$+$16 ($z=0.39$); 
and (d) AC103, AC114, and AC118 ($\bar z=0.31$). 
The dotted lines represent the sliding absolute magnitude cut (see text)
applied for $z_f=5$ ($q_o=0.5, H_o=50$), assuming $M_K = -22.93\,$mag locally.
The solid circles are galaxies whose magnitudes lie
above the magnitude cut, and the open circles represent data falling below
the cut. Least-squares fits to the data above the absolute magnitude
cut, using the slope of a local full FP study by
Pahre et al. (1995), $b=3.2$, are indicated on the plots as solid lines.
\label{figKkorm}}

\figcaption
{$\Delta{\mu^\prime}=\langle\mu\rangle_e-b\log R_e$ 
versus $z$ at standard condition $R_e=1\,$kpc. 
(a) Rest-frame $B$ data with a constant absolute magnitude cut applied, 
$M_B=-19.83\,$mag.
(b) Rest-frame $B$ data with a sliding absolute magnitude cut applied, assuming 
$z_f=5$ and $M_B=-19.83\,$mag locally.
(c) Rest-frame $K$ data with a sliding absolute magnitude cut applied, assuming
$z_f=5$ and $M_K=-22.93\,$mag. The data points are relatively insensitive 
to the choice of $z_f$ and hence are shown only for $z_f=5$.
The curves represent passive evolution predictions from the
1996 solar mass Bruzual \& Charlot models with $q_o=0.5$, $H_o=50$ and
epochs of galaxy formation, $z_f$, as indicated on the figures.
The $X$ symbol represents the values deduced from the local data of
Schade et al.\ (1997) (a, b) and Pahre et al.\ (1995) (c). The other
symbols represent the various cluster data points; their shapes and
shades have no significance. 
\label{Bsc}}
 
\newpage

\begin{deluxetable}{l c c c c c c c c}
\tablecolumns{9}
\tablewidth{0pc}
\tablenum{1}
\tablecaption{Cluster sample and properties \label{tab1}}
\tablehead{
\colhead{Cluster} & \colhead{$z$} &
\multicolumn{3}{c}{T$_{exp}$(ks)} &
\multicolumn{3}{c}{Zero-points (mag)} & \colhead{E(B--V)} \\
\colhead{} & \colhead{}  & \colhead{F555W} & \colhead{F702W}
& \colhead{F814W} & \colhead{F555W} & \colhead{F702W} &
\colhead{F814W} & \colhead{} }
\startdata
A2218     & 0.17 & ---  & 6.5  & ---  & --- &30.774& --- &\phm{$<$} 0.01 \\
A2390     & 0.23 & 8.4  & ---  & 10.5 &30.776& --- &29.892&\phm{$<$} 0.07 \\
AC114     & 0.31 & ---  & 16.6 & ---  & --- &30.988& --- &\phm{$<$} 0.04 \\
AC103     & 0.31 & ---  & 6.5  & ---  & --- &30.774& --- &\phm{$<$} 0.05 \\
Cl1447$+$23 & 0.37 & ---  & 4.2  & ---  & --- &30.727& --- &$<$ 0.03 \\
Cl0024$+$16 & 0.39 & ---  & ---  & 13.2 & --- & --- &29.522&\phm{$<$} 0.03 \\
Cl0303$+$17 & 0.42 & ---  & 12.6 & ---  & --- &30.676& --- &\phm{$<$} 0.12 \\
3C295     & 0.46 & ---  & 12.6 & ---  & --- &30.676& --- &\phm{$<$} 0.00 \\
Cl0412$-$65 & 0.51 & 12.6 & ---  & 14.7 &30.776& --- &29.892&\phm{$<$} 0.00 \\
Cl1601$+$43   & 0.54 & ---  & 16.8 & ---  & --- &30.676& --- &\phm{$<$} 0.00 \\
Cl0016$+$16   & 0.55 & 12.6 & ---  & 16.8 &30.776& --- &29.892&\phm{$<$} 0.03 \\
Cl0054$-$27 & 0.56 & 12.6 & ---  & 16.8 &30.776& --- &29.892&\phm{$<$} 0.00 \\
\enddata
\end{deluxetable}

\begin{deluxetable}{l l l c c c c}
\tablecolumns{7}
\tablewidth{0pc}
\tablenum{2}
\tablecaption{New near-infrared cluster observations \label{tab2}}
\tablehead{
\colhead{} & \colhead{} & \colhead{} &
\colhead{Infrared} & \colhead{Seeing} & \colhead{On Source} &
\colhead{Limiting} \\
\colhead{Cluster} & \colhead{Telescope} & \colhead{Date} &
\colhead{Filter} & \colhead{(arcsec)} & \colhead{Exposure (ks)} &
\colhead{Magnitude}
}
\startdata
Cl0024$+$16&1.5m P60&July 1995&$K_s$&1.4&12.9&18.00 \\
3C295&1.5m P60&July 1995&$K_s$&1.6&15.1&18.25 \\
Cl0412$-$65&3.9m AAT&Nov 1994&$K^\prime$&1.4&3.2&18.50\\
Cl1601$+$43&1.5m P60&July 1995&$K_s$&1.6&15.1&18.25 \\
Cl0016$+$16&1.5m P60&July 1995&$K_s$&1.5&16.5&18.50 \\
Cl0054$-$27&3.9m AAT&Nov 1994&$K^\prime$&1.2&3.6&18.75 \\
\enddata
\end{deluxetable}

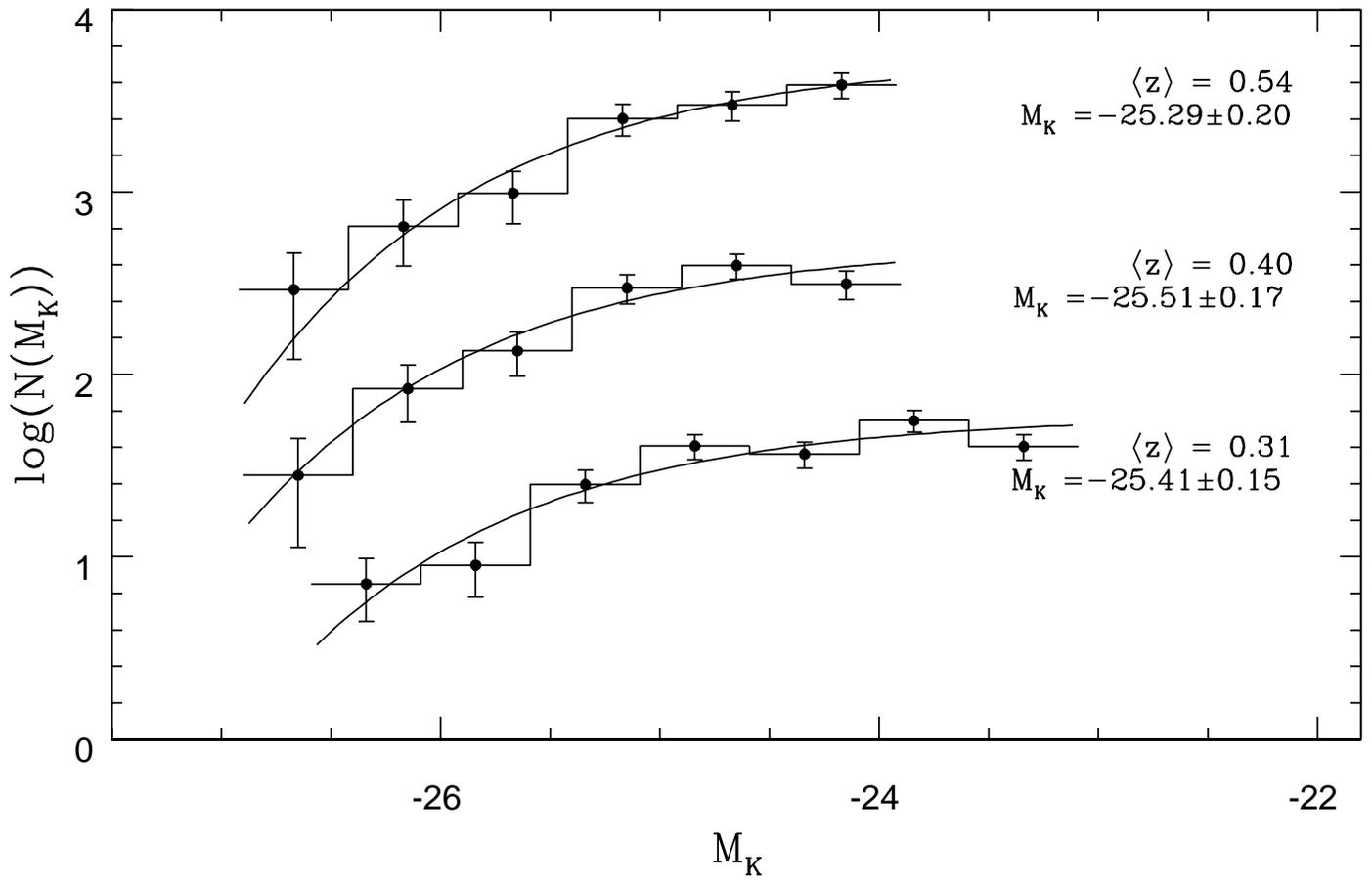
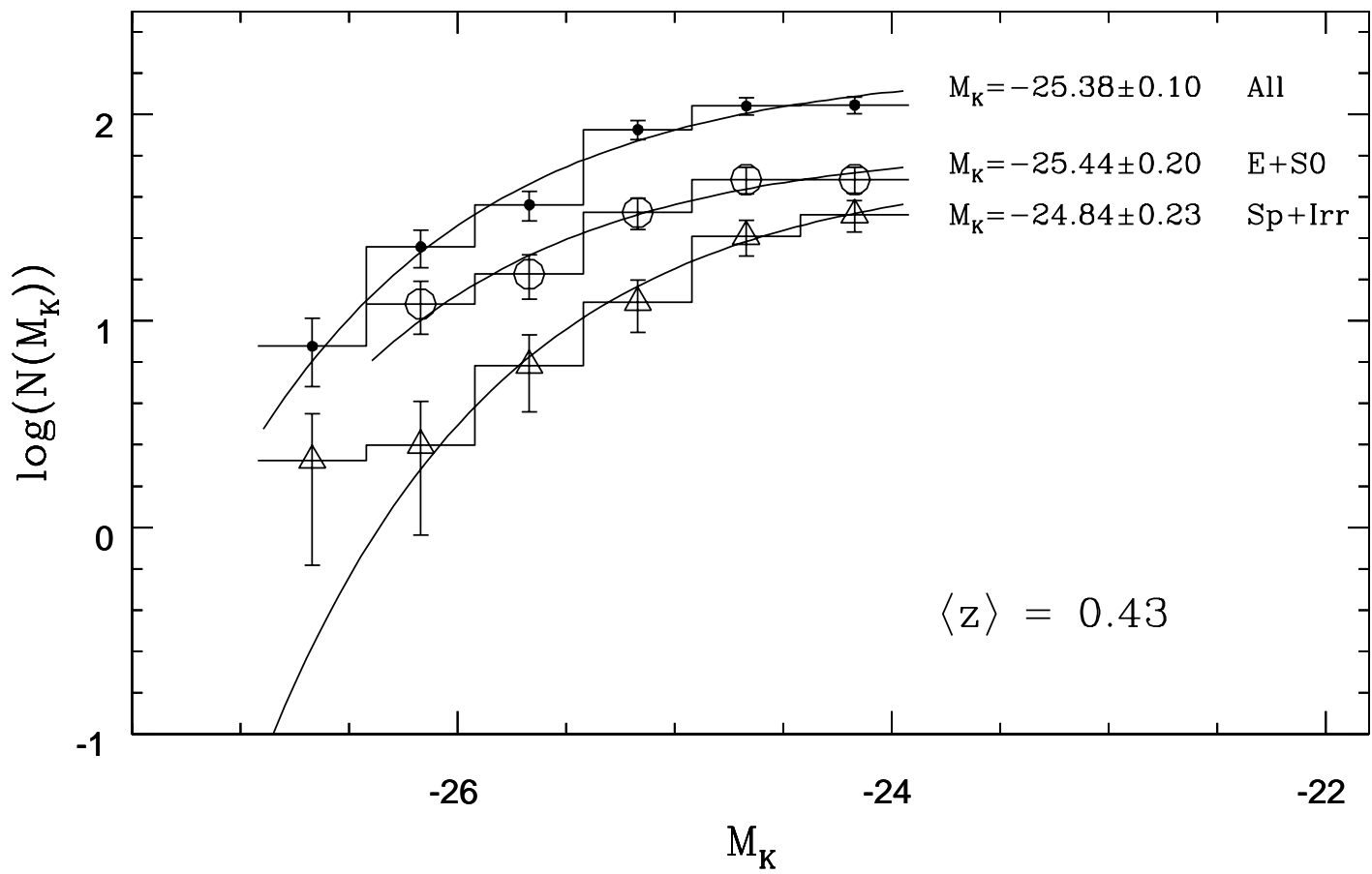
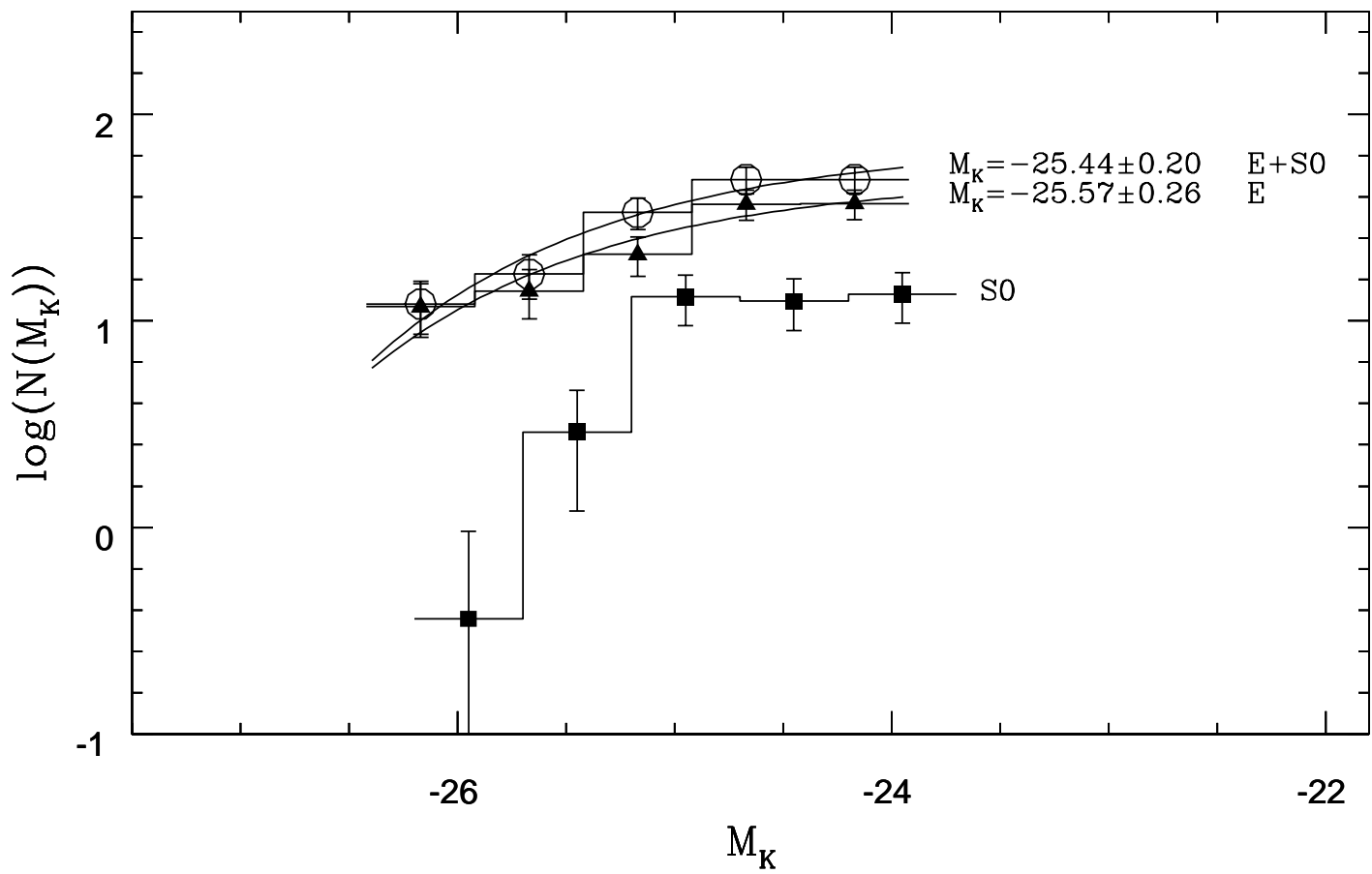
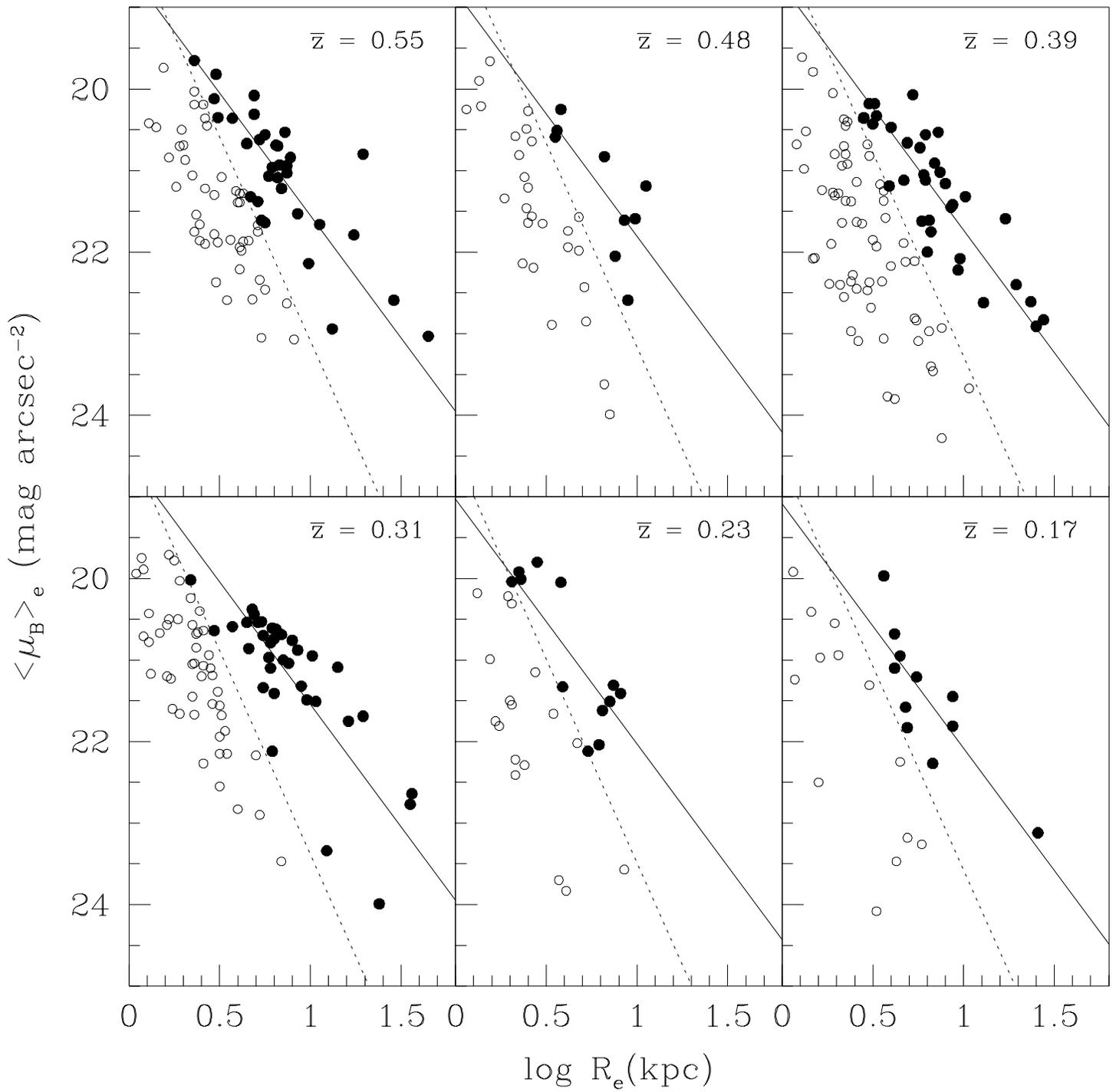
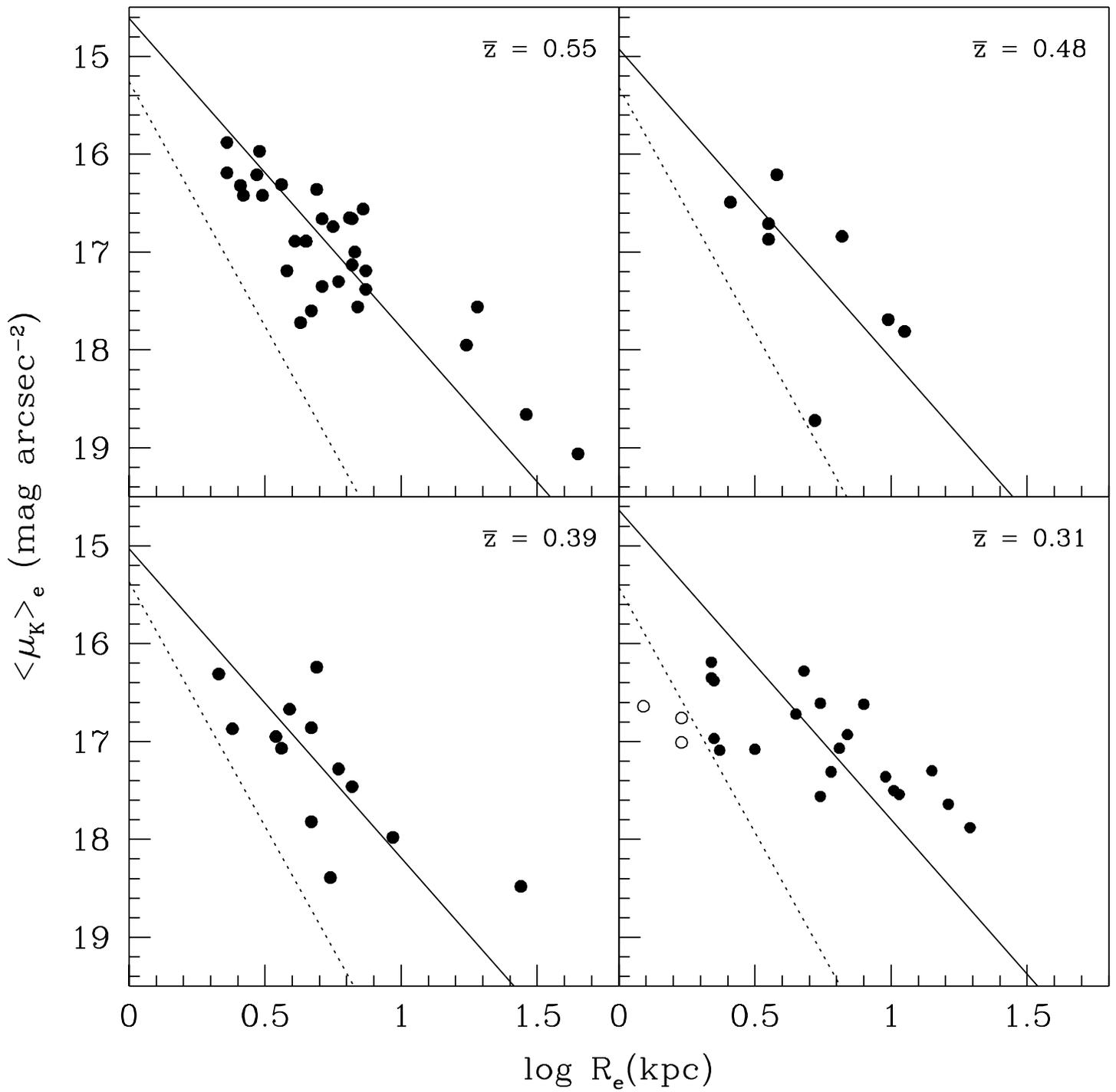
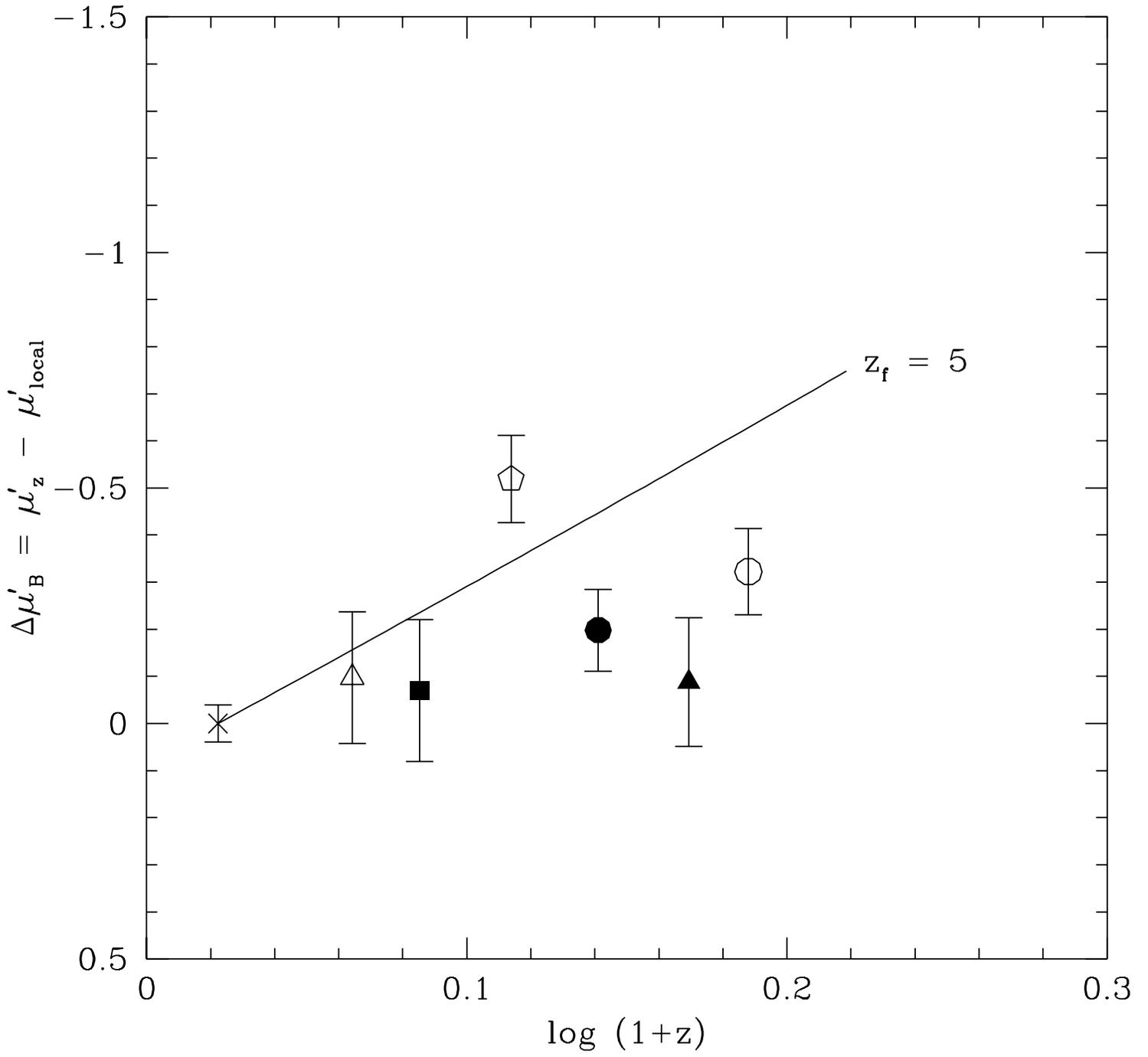
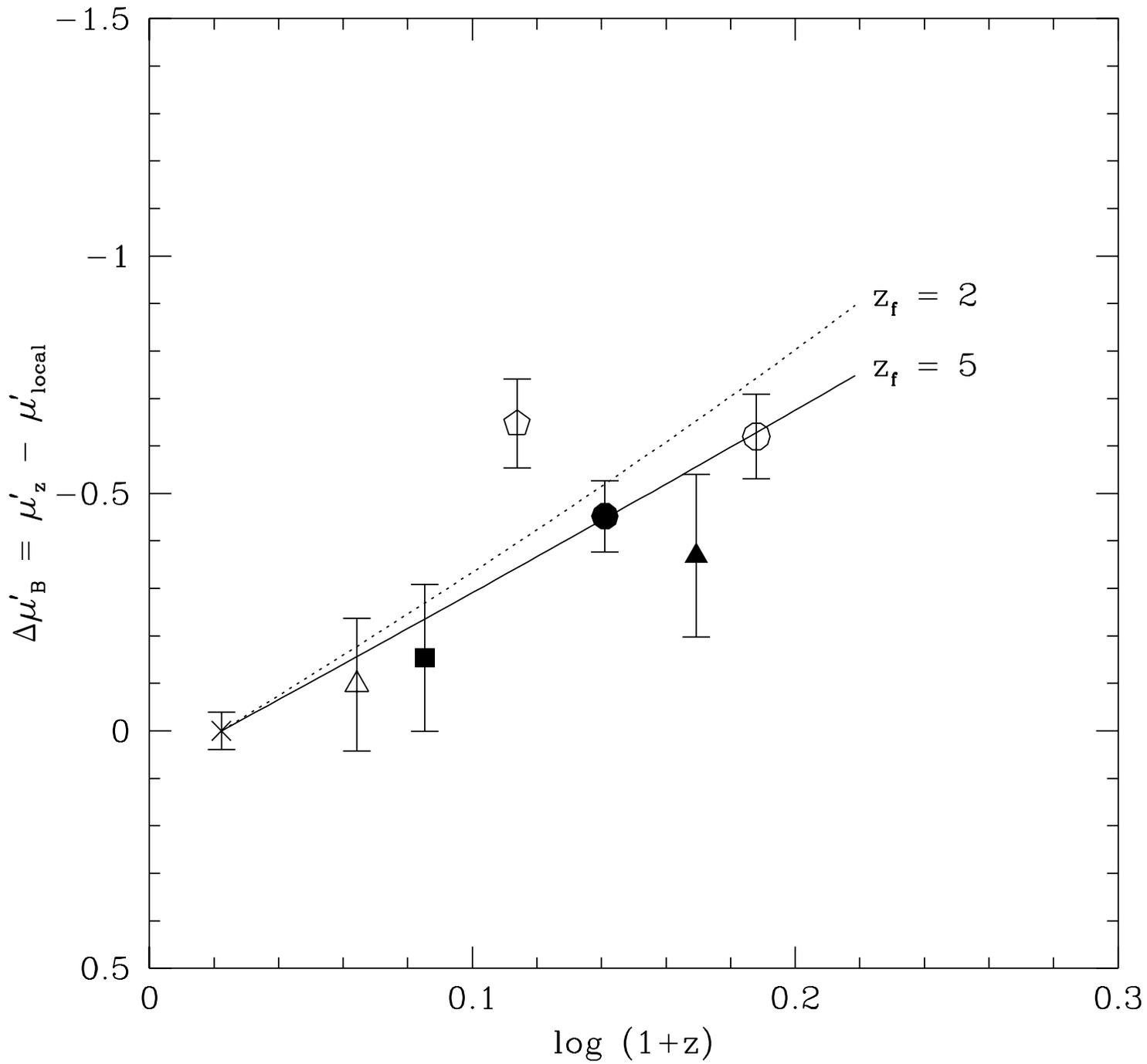
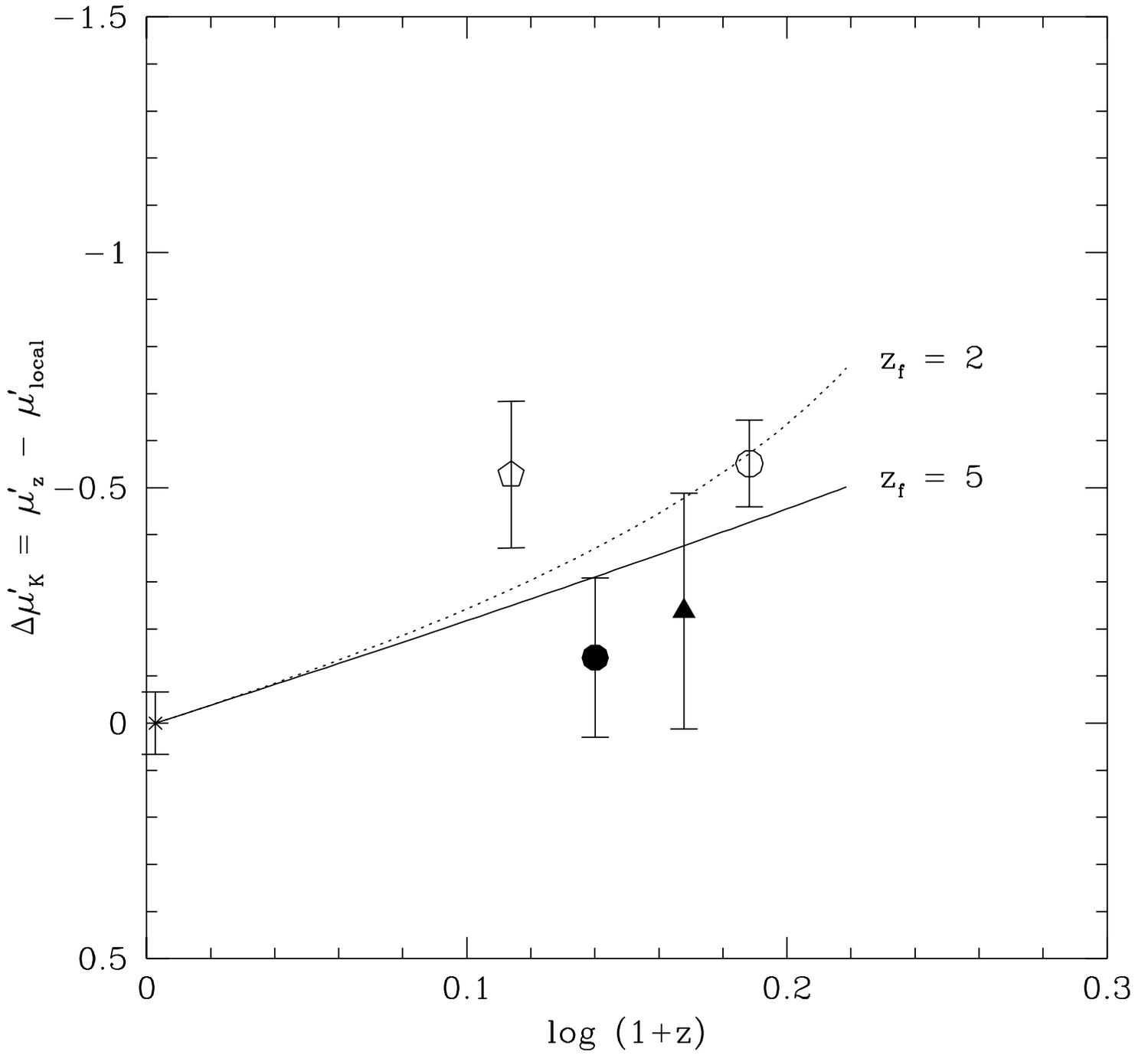
\end{document}